\documentclass[preprint,11pt]{article}
\usepackage{url}

\usepackage{xcolor}
\usepackage{graphicx}
\usepackage{balance}

\usepackage{subfig}                 % 子图包，不要与{subfigure}混用，{subfig}较新
\usepackage{overpic}                % 叭啦叭啦，与图片排版相关
\usepackage{caption}
\usepackage{float}                  % 图片出现在指定位置

\usepackage{bm}                        % 公式加粗
\usepackage{amsthm, amsmath, amssymb}  % 花体字母
\usepackage{mathrsfs}                  % 花体字母

\newtheorem{theorem}{Theorem}
\newtheorem{definition}[theorem]{Definition}
\newtheorem{lemma}[theorem]{Lemma}
\newtheorem{corollary}[theorem]{Corollary}

\def\norm #1{\left\|#1\right\|}

\def\twon #1{\left\|#1\right\|_2}

\def\frobn #1{\left\|#1\right\|_{\text{F}}}

\def\abs #1{\left|#1\right|}

\def\bC{\mathbb{C}}

\def\bR{\mathbb{R}}
\def\bE{\mathbb{E}}

\def\m #1{\boldsymbol{#1}}

\def\cO{\mathcal{O}}

\def\cT{\mathcal{T}}
\def\cU{\mathcal{U}}

\def\bee{\begin{equation}}
\def\ene{\end{equation}}

\def\beq{\begin{eqnarray}}
\def\enq{\end{eqnarray}}
\def\lentwo{\setlength\arraycolsep{2pt}}

\def\equ #1{\begin{equation}#1\end{equation}}
\def\equa #1{\begin{eqnarray}#1\end{eqnarray}}
\def\sbra #1{\left(#1\right)}
\def\mbra #1{\left[#1\right]}
\def\lbra #1{\left\{#1\right\}}
\def\diag #1{\text{diag}#1}

\def\rank#1{\text{rank}#1}

\begin{document}

%\title{Guaranteed Localization and Super-Resolution of More Sources Than Sensors Using Sparse Arrays: Nonasymptotic Analysis of Direct-Augmentation and Spatial-Smoothing ESPRIT}

\title{Nonasymptotic Performance Analysis of Direct-Augmentation and Spatial-Smoothing ESPRIT for Localization of More Sources Than Sensors Using Sparse Arrays}

%\title{Guaranteed Localization and Super-Resolution of More Sources Than Sensors With Finite Snapshots: Nonasymptotic Analysis of Coarray ESPRIT}

\author{Zai Yang and Kaijie Wang\thanks{The authors are with the School of Mathematics and Statistics, Xi'an Jiaotong University, Xi'an 710049, China (e-mail: yangzai@xjtu.edu.cn).}}

%\markboth{Journal of \LaTeX\ Class Files, Vol. 14, No. 8, August 2015}
\maketitle

\begin{abstract}
Direction augmentation (DA) and spatial smoothing (SS), followed by a subspace method such as ESPRIT or MUSIC, are two simple and successful approaches that enable localization of more uncorrelated sources than sensors with a proper sparse array. In this paper, we carry out nonasymptotic performance analyses of DA-ESPRIT and SS-ESPRIT in the practical finite-snapshot regime. We show that their absolute localization errors are bounded from above by $C_1\frac{\max\lbra{\sigma^2, C_2}}{\sqrt{L}}$ with overwhelming probability, where $L$ is the snapshot number, $\sigma^2$ is the Gaussian noise power, and $C_1,C_2$ are constants independent of $L$ and $\sigma^2$, if and only if they can do exact source localization with infinitely many snapshots. We also show that their resolution increases with the snapshot number, without a substantial limit. Numerical results corroborating our analysis are provided.
\end{abstract}

\textbf{Keywords:} Nonasymptotic performance analysis, direct-augmentation ESPRIT, spatial-smoothing ESPRIT, DOA estimation, sparse linear array.

\section{Introduction}
Direction-of-arrival (DOA) estimation refers to the problem of estimating directions of emitting sources from output snapshots of a sensor array \cite{stoica2005spectral}. It is a major topic in array signal processing and has wide applications in radar, sonar, wireless communications, etc \cite{li2007mimo,barbotin2012estimation,lin2007doppler, zeng2018estimation,viberg2018direction,shi2020nested}. It is well-known that up to $N-1$ sources can be resolved by using an $N$-element uniform linear array (ULA). An active research direction is to increase the number of sources resolvable given a fixed number of sensors. This has been done by exploiting rich structures of source signals \cite{viberg2018direction}. A widely studied scenario is when the sources are uncorrelated. In this case, the array covariance matrix of a ULA is Hermitian positive-semidefinite Toeplitz that uniquely determines the source DOAs and powers. Since an $N\times N$ Hermitian Toeplitz matrix is determined by a number $N$ of its entries that can further be captured by an $M\times M$ principal submatrix, where $M$ can be as small as $\cO\sbra{\sqrt{N}}$. This implies that by keeping only $M=\cO\sbra{\sqrt{N}}$ out of $N$ sensors of the ULA, which form an $M$-element sparse linear array (SLA), one may get access to the Toeplitz covariance matrix and estimate up to $\cO\sbra{N}$ sources. This is the principle underlying the fact that a properly designed SLA of $M$ sensors can detect up to $\cO\sbra{M^2}$ sources. Several array geometries of SLAs have been proposed so far, e.g., minimum redundancy arrays (MRAs) \cite{moffet1968minimum,ishiguro1980minimum}, nested arrays \cite{pal2010nested,liu2016super1} and coprime arrays \cite{vaidyanathan2010sparse,qin2015generalized,shi2022enhanced}, by taking into account issues of hardware implementation besides the capacity of source detection.

%Given a SLA, as a subset of a ULA of same aperture, its (difference) coarray reveals which entries of the Toeplitz covariance matrix associated with the ULA can be accessed. If the coarray recovers the aforementioned ULA (consider, e.g., MRAs and nestes arrays), then the SLA is known as a redundancy array and the whole Toeplitz covariance matrix can be recovered from the SLA covariance. Otherwise, there are holes in the coarray and not all of the entries of the Toeplitz covariance can be recovered (consider, e.g., coprime arrays).

%This implies that by keeping only $M=\cO\sbra{\sqrt{N}}$ out of $N$ sensors of the ULA, which form an $M$-element sparse linear array (SLA) and correspond to the $M\times M$ principal submatrix, one may get access to the whole Toeplitz covariance matrix (or a shrunk version of it), known as coarray covariance matrix of the SLA, and estimate up to $\cO\sbra{N}$ sources.

%For example, by using the prior that each emitting source has constant modulus, which is satisfied for , the number of sources resolvable can be doubled \cite{}.

Extensive studies have been made on DOA estimation using SLAs given finitely many snapshots of array output. To enable localization of more sources than sensors, uncorrelated sources are assumed and an existing algorithm usually consists of two steps: 1) construct an augmented covariance matrix given the SLA snapshots or sample covariance matrix and 2) estimate the DOAs from the augmented covariance matrix.
The key to {\em Step 1} lies in how to use the structures of the Toeplitz covariance matrix associated with the virtual ULA containing the SLA. Direct augmentation (DA) \cite{pillai1985new, pillai1987statistical, abramovich1998positive1, abramovich1999positive, abramovich2001detection, liu2015remarks, liu2016super1} is the simplest and a widely used approach that finds the closest Toeplitz approximation, in closed form, to the SLA sample covariance matrix by using the Toeplitz structure. Toeplitz covariance fitting is another common approach that utilizes the Toeplitz and positive semidefinite structures and is usually solved using optimization methods \cite{ottersten1998covariance,abramovich1998positive1,li1999computationally, yang2014discretization,babu2016melt, wu2017toeplitz,zhou2018direction,yang2022robust}. A rank constraint, which reflects the source number, is also included in \cite{babu2016melt,yang2022robust} to further shape the Toeplitz covariance matrix. Given the augmented covariance matrix estimate, {\em Step 2} can usually be accomplished using subspace methods, e.g., MUSIC and ESPRIT. Considering that the covariance estimate produced by DA might not be positive semidefinite, a spatial-smoothing (SS) subspace method has been proposed in {\em Step 2} by formulating the augmented covariance matrix as a single-snapshot output of a virtual, enlarged array \cite{pal2010nested,liu2015remarks}. Sparse estimation methods \cite{yang2018sparse,liu2012direction,tan2014direction,qiao2019guaranteed} have also been combined with such formulation. With the rank constraint, as in \cite{babu2016melt,yang2022robust}, a one-to-one mapping is established between the estimated Toeplitz covariance matrix and the source DOAs and powers by the Carath\'{e}odory-Fej\'{e}r theorem \cite[Theorem 11.5]{yang2018sparse}. Consequently, {\em Step 2} admits a unique solution and the two-step estimation approach essentially becomes a one-step optimization method, leading to higher accuracy and improved robustness to correlated sources \cite{yang2022robust}.

Although many algorithms have been developed for localization of more sources than sensors using SLAs, their theoretical performances are less understood, especially in the regime of finitely many snapshots. Such studies are known as nonasymptotic analysis, as opposed to asymptotic analysis that might not hold in any finite regime (consider, e.g., a great number of asymptotic performance analyses of subspace methods using ULAs \cite{stoica1989music, rao1989performance1, rao1989performance, stoica1990music, stoica1991performance, ottersten1991performance, yuen1996asymptotic, steinwandt2017performance} and SLAs \cite{wang2016coarrays,liu2017cramer}). In this paper, we consider the DA-ESPRIT and (DA-)SS-ESPRIT algorithms that are combinations of the simple DA technique in {\em Step 1} and the prominent ESPRIT and SS-ESPRIT method in {\em Step 2}. We show that the estimation error of DOAs (in fact, the associated spatial frequencies; see details in the main context) of DA-ESPRIT and SS-ESPRIT is bounded from above by $C_1\frac{\max\lbra{\sigma^2,\;C_2}}{\sqrt{L}}$ with overwhelming probability, where $L$ is the snapshot number, $\sigma^2$ is the Gaussian noise power, and $C_1,C_2$ are constants independent of $\sigma^2$ and $L$. This implies that there is no substantial performance gap between the practical scenario of finite $L$ and the limiting case of infinite $L$. The resolution of DOA estimation can be arbitrarily high given $L$ sufficiently large. Our result is derived by quantifying the Toeplitz covariance estimation error using random matrix theory \cite{wainwright2019high}, applying the matrix perturbation theory \cite{stewart1990matrix} and using the latest results on nonasymptotic analysis of ESPRIT \cite{li2020super,yang2023nonasymptotic}. Numerical results corroborating our analyses are also provided.

Our results are related to previous ones in several aspects. Nonasymptotic analysis of ESPRIT using ULAs plays an important role in our analysis. The papers \cite{aubel2016deterministic, fannjiang2016compressive,li2020super} study the single-snapshot ESPRIT with SS preprocessing and show the scaling behavior of the estimation error with respect to the noise level. The paper \cite{yang2023nonasymptotic} investigates the multiple-snapshot ESPRIT and SS-ESPRIT in which the sources can be correlated or coherent and show an error bound proportional to $\frac{\max\lbra{\sigma,\sigma^2}}{\sqrt{L}}$. In contrast to these results, we study the SLA case in this paper that requires different analysis. It is shown that the DOA estimation error has a positive lower bound even in the limiting noiseless case.

Nonasymptotic analysis has also been investigated for DOA estimation using SLAs. Atomic norm or total variation norm methods, which are a kind of gridless sparse methods \cite{yang2018sparse}, have been proposed in \cite{candes2014towards,fernandez2013support,tang2013compressed,tang2014near} for single-snapshot and in \cite{yang2016exact} for multiple-snapshot DOA estimation using ULAs or SLAs. Without the uncorrelated sources assumption, the derived nonasymptotic analyses are usable only when the source number is much smaller than the sensor number. Moreover, the DOAs need to be sufficiently separated. Similar results have been derived in \cite{liao2014music} for MUSIC without the uncorrelated sources assumption. Uncorrelated sources are assumed in \cite{tan2014direction,qiao2019guaranteed} to enable localization of more sources than sensors. The paper \cite{tan2014direction} proposed a DA plus single-snapshot atomic norm method for DOA estimation using SLAs. By applying the results in \cite{candes2014towards,fernandez2013support}, it is shown that more sources than sensors can be resolved with a separation condition weaker than that in \cite{yang2016exact}. Similar results have been derived in \cite{qiao2019guaranteed} by assuming on-grid DOAs, a common assumption in compressed sensing or sparse methods \cite{yang2018sparse}, with a DA plus $\ell_1$ optimization method. Differently from previous nonasymptotic analyses, we consider the simple DA-ESPRIT and SS-ESPRIT algorithms and show that the separation condition can be removed, implying that DA-ESPRIT and SS-ESPRIT have higher resolution, at least in the presence of a large number of snapshots.

Notations used in this paper are as follows. The set of real and complex numbers are denoted by $\bR$ and $\bC$ respectively. Boldface letters are reserved for vectors and matrices. For integer $N$, we define the set $[N] = \lbra{0,\dots,N-1}$. The amplitude of scalar $a$ is denoted by $\abs{a}$. The transpose, complex transpose and pseudo-inverse of matrix $\m{A}$ are denoted by  $\m{A}^T$, $\m{A}^H$ and $\m{A}^{\dag}$ respectively. The rank, spectral norm and Frobenius norm of matrix $\m{A}$ are denoted by $\rank\sbra{\m{A}}$, $\norm{\m{A}}$ and $\frobn{\m{A}}$. The $j$th greatest eigenvalue (or singular value) of a matrix is denoted by $\lambda_j\sbra{\cdot}$ (or $\sigma_j\sbra{\cdot}$). The $j$th entry of vector $\m{x}$ is $x_j$, and the $(j,l)$ entry of matrix $\m{A}$ is $A_{jl}$. For vector $\m{x}$, $\diag\sbra{\m{x}}$ denotes a diagonal matrix with $\m{x}$ on the diagonal, and $\twon{\m{x}}$ denotes the Euclidean norm. The expectation of a random variable is denoted $\bE[\cdot]$.

The rest of the paper is organized as follows. Section \ref{sec:algorithm} introduces the problem formulation of DOA estimation using SLAs and the DA-ESPRIT and SS-ESPRIT algorithms. Section \ref{sec:analysis} presents our asymptotic analyses of DA-ESPRIT and SS-ESPRIT. Section \ref{sec:simulation} provides numerical results validating our analyses. Section \ref{sec:conclusion} concludes the paper.

\section{DA-ESPRIT and SS-ESPRIT for DOA Estimation Using SLAs} \label{sec:algorithm}

\subsection{Problem Formulation}
Consider an $N_{\text{S}}$-element SLA that is a subset of a virtual $N$-element ULA and is denoted by the index set $\Omega=\lbra{\Omega_1,...,\Omega_{N_{\text{S}}}}$, where $0 \le \Omega_1<...<\Omega_{N_{\text{S}}} \le N-1$. It becomes the ULA in the special case $N_{\text{S}} = N$. Assume that $K$ narrowband, far-field sources impinge on the array from distinct directions $\theta_{k}\in\left[ -90^{\circ},90^{\circ} \right), k=1,\dots,K$, and the elements of the virtual ULA are spaced by half a wavelength apart. Then, in the case $N_{\text{S}} = N$, the $L$-snapshot sensor output can be modeled as \cite{stoica2005spectral}:
\begin{equation} \label{eq:ULAmodel}
	\m{Y} = \m{A}\m{S} + \m{E},
\end{equation}
where $\m{Y}$ is an $N \times L$ output matrix, $\m{S}$ is a $K \times L$ source signal matrix, $\m{E}$ is an $N \times L$ noise matrix (note that each column of $\m{Y},\m{S},\m{E}$ corresponds to one snapshot), and $\m{A}$ is an $N \times K$ array manifold matrix whose $\sbra{n,k}$ entry is given by $e^{i\pi\sbra{n-1}\sin\theta_k}$ with $i=\sqrt{-1}$. In the general SLA case, the $L$-snapshot output is given by
\begin{equation} \label{Y_Omega}
	\m{Y}_{\Omega} = \m{A}_{\Omega}\m{S} + \m{E}_{\Omega},
\end{equation}
where $\m{Y}_{\Omega}$, $\m{A}_{\Omega}$ and $\m{E}_{\Omega}$ are the submatrices of $\m{Y}$, $\m{A}$ and $\m{E}$ by retaining the rows indexed by $\Omega+1\triangleq \lbra{\Omega_j+1}$, respectively (note that the term ``$+1$'' appears since the elements of $\Omega$ start with 0 and the row/column index of a matrix starts with 1). Note that the $k$th DOA $\theta_k$ has a one-to-one relation to the spatial frequency $f_k=\frac{1}{2}\sin\theta_k \ \text{mod} \ 1\in\left[ 0,1 \right)$. Consequently, we study estimation of $\lbra{f_k}$ hereafter for convenience. This transforms the DOA estimation problem as spatial spectral analysis, where $\lbra{f_k}$ play as frequency parameters in the data model. Our objective is to estimate $\lbra{f_k}$ given $\m{Y}_{\Omega}$ and $K$.

We will make the following assumptions throughout this
paper:
\begin{itemize}
	\item[\textbf{A1:}] The columns of $\m{S}$ are i.i.d.~circularly symmetric complex Gaussian with zero mean and diagonal covariance $\m{\Sigma}=\diag\sbra{\m{p}}$, where $\m{p}=\mbra{p_1,\dots,p_K}^T$ with all $p_k$'s positive;
	\item[\textbf{A2:}] The entries of $\m{E}$ are i.i.d.~circularly symmetric complex Gaussian with zero mean and variance $\sigma^2$;
	\item[\textbf{A3:}] $\m{S}$ and $\m{E}$ are independent.
\end{itemize}

\subsection{DA-ESPRIT for DOA Estimation Using SLAs}
We first consider the special $N$-element ULA case. In this case and under assumptions A1--A3, the array output at each snapshot (one column of $\m{Y}$ in \eqref{eq:ULAmodel}) is circularly symmetric complex Gaussian distributed with zero mean and Toeplitz variance
\begin{equation}\begin{split}
\m{R}= \bE\mbra{\dfrac{1}{L}\m{Y}\m{Y}^H} = \m{A}\m{\Sigma}\m{A}^H + \sigma^2\m{I}= \begin{bmatrix}
		r_0 & r_{-1} & \cdots & r_{1-N} \\
		r_1 & r_0 & \cdots & r_{2-N} \\
		\vdots & \vdots & \ddots& \vdots \\
		r_{N-1} & r_{N-2} & \cdots & r_0
	\end{bmatrix},
\end{split} \label{eq:R}
\end{equation}
where
\equ{r_j = \left\{\begin{array}{ll} \sum_{k=1}^K p_k+\sigma^2, & \text{ if } j=0,\\ \sum_{k=1}^K p_k e^{i2\pi j f_k}, & \text{ otherwise.} \end{array} \right. \label{eq:rj}}
Note that the Toeplitz covariance $\m{R}$ uniquely determines the source DOAs and powers $\lbra{f_k,p_k}_{k=1}^K$ and the noise power $\sigma^2$ by the Carath\'{e}odory-Fej\'{e}r theorem \cite[Theorem 11.5]{yang2018sparse}, that can be computed by the ESPRIT algorithm detailed below.

In practice, we have only an estimate of $\m{R}$, denoted by $\widehat{\m{R}}$. The ESPRIT algorithm estimates the frequencies $\lbra{f_k}$ from $\widehat{\m{R}}$. First, we compute the eigen-decomposition of $\widehat{\m{R}}$:
\equ{\widehat{\m{R}} = \sum_{j=1}^{N} \widehat{\lambda}_j \widehat{\m{u}}_j\widehat{\m{u}}_j^H = \widehat{\m{U}} \widehat{\m{\Lambda}} \widehat{\m{U}}^H + \widehat{\m{U}}_{\perp} \widehat{\m{\Lambda}}_{\perp} \widehat{\m{U}}_{\perp}^H,}
where $\lbra{\widehat{\lambda}_j}$ are eigenvalues sorted in descending order, $\widehat{\m{\Lambda}}$ is a diagonal matrix with the greatest $K$ eigenvalues $\widehat{\lambda}_1,\dots,\widehat{\lambda}_K$ on the diagonal, $\widehat{\m{U}}$ is composed of the associated $K$ eigenvectors $\widehat{\m{u}}_1,\dots,\widehat{\m{u}}_K$, and $\widehat{\m{\Lambda}}_{\perp}, \widehat{\m{U}}_{\perp}$ are composed by the other eigenvalues and eigenvectors. Then, let $\widehat{\m{U}}_1,\widehat{\m{U}}_2$ be submatrices of $\widehat{\m{U}}$ obtained by removing the last and the first rows, respectively, and we compute the eigenvalues $\lbra{\widehat{z}_k}$ of $\widehat{\m{U}}_1^{\dag}\widehat{\m{U}}_2$. Finally, we obtain the frequency estimates as
\begin{equation}
	\widehat{f}_k = \frac{1}{2\pi}\arg{\frac{\widehat{z}_k}{\abs{\widehat{z}_k}}}, \quad k = 1,\dots,K.
\end{equation}
Suppose that $\widehat{\m{R}}$ is given by its ground truth $\m{R}$. It can then be verified that the signal subspace $\widehat{\m{U}}$ equals the ground truth $\m{U}$ that is identical to the range space of $\m{A}$ and $\widehat{f}_k$ equals the true value $f_k$.

We now consider DOA estimation using the $N_\text{S}$-element SLA $\Omega$. In this case and under assumptions A1--A3, the array output covariance matrix $\m{R}_{\Omega} =\bE\mbra{\dfrac{1}{L}\m{Y}_{\Omega} \m{Y}_{\Omega} ^H}$ is a principal submatrix of $\m{R}$ by taking its rows and columns indexed by the set $\Omega+1$. It can easily been shown that we have access to those $r_j$'s in \eqref{eq:R} indexed by the difference set of $\Omega$,
\equ{\Omega_{\text{ca}} = \lbra{\Omega_j - \Omega_l:\; 1\leq l \leq j \leq N_{\text{S}}},}
that is referred to as the coarray of the SLA $\Omega$. Let $M$ be the greatest integer such that the ULA $[M]$ is a subset of the coarray $\Omega_{\text{ca}}$, where $M = \cO\sbra{N_{\text{S}}^2}$ for well-designed SLAs. It follows that we have full access to the $M\times M$ leading principal submatrix matrix of $\m{R}$, $\m{R}_{[M]}$, that corresponds to the output covariance matrix of the ULA $[M]$. This means that we have augmented the covariance matrix from $\m{R}_{\Omega}$ to $\m{R}_{[M]}$. Given $\m{R}_{[M]}$, up to $\cO\sbra{N_{\text{S}}^2}$ sources can be exactly localized using ESPRIT.

In practice, the SLA output covariance matrix $\m{R}_{\Omega}$ is estimated as the sample covariance matrix
\equ{\widehat{\m{R}}_{\Omega} =\dfrac{1}{L}\m{Y}_{\Omega} \m{Y}_{\Omega}^H.}
What DA does is to simply estimate $\lbra{r_m}_{m=0}^{M-1}$ in \eqref{eq:R} as the mean of the associated entries in $\widehat{\m{R}}_{\Omega}$:
\equ{\widehat{r}_m = \frac{\sum_{(j,l):\; \Omega_j - \Omega_l = m} \mbra{\widehat{R}_{\Omega}}_{jl}} {\sum_{(j,l):\; \Omega_j - \Omega_l = m} 1}, }
where $1\leq j, l\leq N_{\text{S}}$. By using these covariance lags, we are now able to construct the Toeplitz estimate of $\m{R}_{[M]}$, denoted by
\equ{\widehat{\m{R}}_{\text{DA}} \triangleq \widehat{\m{R}}_{[M]} = \begin{bmatrix}
		\widehat{r}_0 & \widehat{r}_{-1} & \cdots & \widehat{r}_{1-M} \\
		\widehat{r}_1 & \widehat{r}_0 & \cdots & \widehat{r}_{2-M} \\
		\vdots & \vdots & \ddots& \vdots \\
		\widehat{r}_{M-1} & \widehat{r}_{M-2} & \cdots & \widehat{r}_0
	\end{bmatrix}. \label{eq:RDA}}
DA-ESPRIT estimates the frequencies by applying ESPRIT to the DA covariance matrix estimate $\widehat{\m{R}}_{\text{DA}}$.

It is worth noting that as $L$ approaches infinity, the sample covariance matrix $\widehat{\m{R}}_{\Omega}$ approaches $\m{R}_{\Omega}$ and $\widehat{\m{R}}_{\text{DA}}$ approaches its ground truth
\equ{\m{R}_{\text{DA}} = \m{R}_{[M]} = \m{A}_M\m{\Sigma}\m{A}_M^H + \sigma^2\m{I}, \label{eq:RDA2}}
where $\m{A}_M$ is the Vandermonde matrix composed of the first $M$ rows of $\m{A}$ and the second equality follows from \eqref{eq:R}. Therefore, DA-ESPRIT is statistically consistent in the snapshot number $L$ regardless of the DOAs given the source number $K\leq M-1$.

\subsection{SS-ESPRIT for DOA Estimation Using SLAs}
SS-ESPRIT has been used in \cite{pal2010nested,liu2015remarks} to resolve the problem that the DA covariance estimate $\widehat{\m{R}}_{\text{DA}}$ in \eqref{eq:RDA} might not be positive semidefinite. Observe by \eqref{eq:rj} that the length-$(2M-1)$ vector $\m{r} \triangleq\mbra{r_{1-M}, \dots, r_{-1}, r_0, r_1,\dots,r_{M-1}}^T$ is a single-snapshot output of a $(2M-1)$-element ULA, and thus SS-ESPRIT can be used to do DOA estimation based on $\widehat{\m{r}}$ (note that the presence of $\sigma^2$ in $r_0$ does not change the final DOA estimates). The SS covariance estimate is given by \cite{pal2010nested}
\equ{\widehat{\m{R}}_{\text{SS}} = \frac{1}{M}\widehat{\m{R}}_{[M]} \widehat{\m{R}}_{[M]}^H = \frac{1}{M}\widehat{\m{R}}_{[M]}^2 = \frac{1}{M} \widehat{\m{R}}_{\text{DA}}^2. \label{eq:RSSDA}}
It can easily be shown that SS-ESPRIT is equivalent to DA-ESPRIT in the asymptotic setup, implying that SS-ESPRIT is also statistically consistent regardless of the DOAs given $K\leq M-1$. But this might not be true with finitely many snapshots, as detailed in the ensuing section.

\section{Nonasymptotic Analyses of DA-ESPRIT and SS-ESPRIT} \label{sec:analysis}
We carry out nonasymptotic performance analyses for DA-ESPRIT and SS-ESPRIT in this section. The error of DOA estimation using DA-ESPRIT or SS-ESPRIT originates from the estimation error of the DA covariance matrix. To analyze the DOA or frequency estimation error, we first introduce the recent nonasymptotic analysis of ESPRIT that relates the frequency estimation error to the signal subspace estimation error of ESPRIT. We then quantify the latter error by analyzing the DA covariance estimation error and applying the matrix perturbation theory. The final error bound is obtained by combining these results.

%The error of DOA estimation using subspace methods originates from the error between the true and estimated signal subspaces. In this section, to bound the frequency estimation error of DA- and SS-ESPRIT, we will first introduce the metric of DOA estimation error using ESPRIT, the metric of distance of subspaces, and their relationship. Then, we derive a tighter signal subspace perturbation bound for DA and SS, which utilizes the Davis-Kahan $\sin\Theta$ theorem and the randomness of the signals and noise. Finally, combining the perturbation bound and the relationship between frequency estimation error and signal subspace error, we derive the error bound for DA- and SS-ESPRIT.

\subsection{Useful Results}

Consider two $r$-dimensional linear subspaces $\cU$, $\widehat{\cU}$ in $\bC^p$ associated with the column spaces of $p \times r$ isometric matrices $\m{U}$, $\widehat{\m{U}}$, where $\m{U}^H\m{U} = \widehat{\m{U}}^H\widehat{\m{U}} = \m{I}$. We will not distinguish a subspace $\cU$ and its matrix representation $\m{U}$ hereafter whenever it is clear from the context. The canonical angles between $\m{U}$ and $\widehat{\m{U}}$ are defined as \equ{\theta_j\sbra{\widehat{\m{U}},\m{U}} = \arccos \sigma_j\sbra{\widehat{\m{U}}^H \m{U}},\quad j=1,\dots,r,}
where $\sigma_j$ denotes the $j$th greatest singular value.
Define matrices
\equ{\Theta = \begin{bmatrix}\theta_1 & & \\ & \ddots & \\ & & \theta_r \end{bmatrix},\quad \sin\Theta = \begin{bmatrix}\sin\theta_1 & & \\ & \ddots & \\ & & \sin\theta_r \end{bmatrix}.}
The distance between $\m{U}$ and $\widehat{\m{U}}$ can be defined as \cite{chen2021spectral}
\equ{\text{dist}\sbra{\widehat{\m{U}},\m{U}} = \norm{\sin\Theta\sbra{\widehat{\m{U}},\m{U}}}. \label{eq:distdef}}

For the frequency set $\cT = \lbra{f_k}_{k=1}^K$ and its estimate $\widehat{\cT} = \lbra{\widehat{f}_k}_{k=1}^K$, the matched distance between $\cT$ and $\widehat{\cT}$ measures the maximum absolute wrap-around error of frequency estimation and is defined as \cite{li2022stability,yang2023nonasymptotic}:
\begin{equation}
	\begin{split}
\text{md} \sbra{\widehat{\cT},\cT} = \min_{\psi}\max_k\min\lbra{\abs{\widehat{f}_{\psi\sbra{k}}-f_k} , 1-\abs{\widehat{f}_{\psi\sbra{k}}-f_k}},
	\end{split}
\end{equation}
where $\psi$ is a permutation on $\lbra{1,\cdots,K}$. The following lemma \cite[Lemma \uppercase\expandafter{\romannumeral10}.2]{li2022stability} shows that for ESPIRT the matched distance can be bounded from above by the distance between $\m{U}$ and $\widehat{\m{U}}$.
\begin{lemma}
Let $\widehat{\m{U}}$ be an estimate of the signal subspace with an $N$-element ULA, where $N\geq K+1$. Then, it holds for ESPRIT that
	\equ{\text{md}\sbra{\widehat{\cT},\cT} \leq \frac{2^{2K+4}\sqrt{K^3N}} {\sigma_K\sbra{\m{A}}} \cdot \text{dist}\sbra{\widehat{\m{U}},\m{U}}. \label{eq:mdlem}} \label{lem:mdlem}
\end{lemma}

Weyl's inequality \cite[Theorem 4.3.1]{horn2012matrix} and the Davis-Kahan $\sin\Theta$ theorem \cite{davis1970rotation}, \cite[Corollary 2.8]{chen2021spectral} are well-known results in matrix perturbation theory that measure perturbations of eigenvalues and subspaces, respectively, and are given in the following theorem.
\begin{theorem} Consider $p\times p$ Hermitian matrices $\m{R}$ and $\widehat{\m{R}}=\m{R}+\m{E}$ that admit the eigen-decompositions:
	{\lentwo\equa{ \m{R}
		&=& \sum_{j=1}^p \lambda_j \m{u}_j\m{u}_j^H = \begin{bmatrix}\m{U} & \m{U}_{\perp} \end{bmatrix} \begin{bmatrix}\m{\Lambda} & \\ & \m{\Lambda}_{\perp} \end{bmatrix} \begin{bmatrix}\m{U}^H \\ \m{U}_{\perp}^H \end{bmatrix}, \\ \widehat{\m{R}}
		&=& \sum_{j=1}^p \widehat{\lambda}_j \widehat{\m{u}}_j\widehat{\m{u}}_j^H = \begin{bmatrix}\widehat{\m{U}} & \widehat{\m{U}}_{\perp} \end{bmatrix} \begin{bmatrix}\widehat{\m{\Lambda}} & \\ & \widehat{\m{\Lambda}}_{\perp} \end{bmatrix} \begin{bmatrix}\widehat{\m{U}}^H \\ \widehat{\m{U}}_{\perp}^H \end{bmatrix},
	}}where the eigenvalues $\lbra{\lambda_j}$ and $\lbra{\widehat{\lambda}_j}$ are sorted in descending order, and $\m{U}$, $\m{\Lambda}$, $\widehat{\m{U}}$, $\widehat{\m{\Lambda}}$ are composed of the first $r<p$ eigenvectors or eigenvalues. Then, it holds that
\begin{equation}
\abs{\widehat{\lambda}_j-\lambda_j} \le \norm{\m{E}}, \quad j=1,\dots,p.
\end{equation}
If $\norm{\m{E}}\leq 0.293\sbra{\lambda_r - \lambda_{r+1}}$, then
\equ{\text{dist}\sbra{\widehat{\m{U}},\m{U}} \leq \frac{2\norm{\m{E}}}{\lambda_r - \lambda_{r+1}}.} \label{thm:DK}
\end{theorem}

The following result quantifies the error of Gaussian covariance estimation and is referred to \cite[Example 6.3]{wainwright2019high}.
\begin{lemma} Let each column of $p\times n$ matrix $\m{X}$ be i.i.d.~sampled from a circularly symmetric complex Gaussian distribution with zero mean and covariance $\m{\Sigma}$. Then, it holds that
\equ{\norm{\frac{1}{n} \m{X}\m{X}^H - \m{\Sigma}} \leq \lbra{2\sbra{\sqrt{\frac{p}{n}}+u} + \sbra{\sqrt{\frac{p}{n}}+u}^2}\norm{\m{\Sigma}}}
with probability at least $1-2e^{-\frac{n}{2}u^2}$. \label{lem:GaussCovErr}
\end{lemma}

%Besides, considering DA and SS may yield different signal subspace estimations, which is caused by the non-positive definiteness of $\widehat{\m{R}}_{\text{DA}}$, we will use a consequence of a more general result known as Weyl's inequality \cite[Theorem 4.3.1]{horn2012matrix} to restrict the positive definiteness of $\widehat{\m{R}}_{\text{DA}}$.
%\begin{lemma} \label{lem:Weyl}
%	Consider $p\times p$ Hermitian matrices $\m{M}$, $\widehat{\m{M}}=\m{M}+\m{E}$, and their eignvalues $\lbra{\lambda_j}$, $\lbra{\widehat{\lambda}_j}$ that are sorted in descending order, then it holds that
%	\begin{equation}
%		\max_{j}\abs{\widehat{\lambda}_j-\lambda_j} \le \norm{\m{E}}.
%	\end{equation}
%\end{lemma}

\subsection{Signal Subspace Estimation Error of DA-ESPRIT}
DA-ESPRIT obtains the signal subspace estimate $\widehat{\m{U}}$ from the eigen-decomposition of $\widehat{\m{R}}_{\text{DA}}$, while its ground truth $\m{U}$ is associated with $\m{R}_{\text{DA}}$ in \eqref{eq:RDA2}. The following lemma relates the DA covariance estimation error to the error of the SLA sample covariance matrix.

\begin{lemma} It holds that
\begin{equation}
		\begin{split}
			\norm{\widehat{\m{R}}_{\text{DA}}-\m{R}_{\text{DA}}}
			 \le \sqrt{MN_{\text{S}}} \norm{\widehat{\m{R}}_{\Omega}-\m{R}_{\Omega}}.
		\end{split}
	\end{equation} \label{lem:RDAerr}
\end{lemma}
\begin{proof}Observe that each entry of $\m{r}$ appears in $\m{R}_{\text{DA}}$ at most $M$ times and appears in $\m{R}_{\Omega}$ at least once. Thus, we have
\begin{equation}
		\begin{split}
			\norm{\widehat{\m{R}}_{\text{DA}}-\m{R}_{\text{DA}}}
			& \le \frobn{\widehat{\m{R}}_{\text{DA}}-\m{R}_{\text{DA}}} \\
			& \le \sqrt{M} \twon{\widehat{\m{r}}-\m{r}} \\
			& \leq \sqrt{M}  \frobn{\widehat{\m{R}}_{\Omega} - \m{R}_{\Omega}} \\
			& \le \sqrt{MN_{\text{S}}} \norm{\widehat{\m{R}}_{\Omega}-\m{R}_{\Omega}},
		\end{split}
	\end{equation}
completing the proof.
\end{proof}

Let $p_{\max} = \max_k p_k$ and $p_{\min} = \min_k p_k$. We have the following result.

\begin{lemma} Under Assumptions A1-A3, if $L\geq N_S$, then
\equ{\norm{\widehat{\m{R}}_{\Omega}-\m{R}_{\Omega}} \leq 8\sqrt{\frac{N_S}{L}}\sbra{p_{\max}\norm{\m{A}_{\Omega}}^2+\sigma^2 }. }
with probability at least $1-2e^{-\frac{N_S}{2}}$. \label{lem:ROmegaerr}
\end{lemma}
\begin{proof} Under Assumptions A1-A3, all columns of $\m{Y}_{\Omega}$ are i.i.d.~Gaussian with zero mean and covariance $\m{R}_{\Omega}$, and $\widehat{\m{R}}_{\Omega}$ is an estimate of the Gaussian covariance $\m{R}_{\Omega}$. Applying Lemma \ref{lem:GaussCovErr} and letting $u = \sqrt{\frac{N_S}{L}}$, we obtain
\equ{\norm{\widehat{\m{R}}_{\Omega} - \m{R}_{\Omega}} \leq 4\sbra{\sqrt{\frac{N_S}{L}} + \frac{N_S}{L}}\norm{\m{R}_{\Omega}} \leq 8\sqrt{\frac{N_S}{L}}\norm{\m{R}_{\Omega}}}
with probability at least $1-2e^{-\frac{N_S}{2}}$, which further concludes the lemma by noting that
\equ{\begin{split}\norm{\m{R}_{\Omega}}
= \norm{\m{A}_{\Omega}\m{\Sigma}\m{A}_{\Omega}^H + \sigma^2\m{I}} \leq \norm{\m{A}_{\Omega}}^2 \norm{\m{\Sigma}} + \sigma^2 = p_{\max}\norm{\m{A}_{\Omega}}^2 + \sigma^2. \end{split}}
\end{proof}

\begin{theorem} \label{thm:DASubspaceError}
Let $\widehat{\m{U}}$ be the estimated signal subspace associated with $\widehat{\m{R}}_{\text{DA}}$ and $\m{U}$ be its ground truth. Under Assumptions A1-A3, if $M \ge K+1$, $L\geq N_S$ and the upper bound below is less than $0.586$, then it holds that
	\begin{equation} \label{eq:DASubspaceError}
\text{dist}\sbra{\widehat{\m{U}}, \m{U}}  \le \frac{16N_S\sqrt{M}}{p_{\min}\sigma^2_K\sbra{\m{A}_M}} \cdot \frac{p_{\max}\norm{\m{A}_{\Omega}}^2 +\sigma^2}{\sqrt{L}}.
	\end{equation}
	with probability at least $1 - 2e^{-\frac{N_{\text{S}}}{2}}$.
\end{theorem}

\begin{proof}
	We apply Theorem \ref{thm:DK} to bound the distance between $\m{U} $ and $ \widehat{\m{U}}$:
	\begin{equation} \label{eq:DASubspaceBoundApplyingDavis-Kahan}
		\text{dist} \sbra{\widehat{\m{U}},\m{U}} \le \frac{2\norm{\widehat{\m{R}}_{\text{DA}}-\m{R}_{\text{DA}}}} {\lambda_K\sbra{\m{R}_{\text{DA}}}-\lambda_{K+1}\sbra{\m{R}_{\text{DA}}}}.
	\end{equation}
By combining Lemma \ref{lem:RDAerr} and Lemma \ref{lem:ROmegaerr}, we have
\begin{equation}
\begin{split}
			\norm{\widehat{\m{R}}_{\text{DA}}-\m{R}_{\text{DA}}}
			 \le 8N_S\sqrt{\frac{M}{L}}\sbra{p_{\max} \norm{\m{A}_{\Omega}}^2+\sigma^2 }
		\end{split} \label{eq:RDAerr2}
\end{equation}
with probability at least $1-2e^{-\frac{N_S}{2}}$.
Moreover, it follows from \eqref{eq:RDA2} that
	\begin{equation}
		\begin{split}
			\lambda_j\sbra{\m{R}_{\text{DA}}} = \lambda_j\sbra{\m{A}_M\m{\Sigma}\m{A}_M^H + \sigma^2\m{I}}  = \lambda_j\sbra{\m{A}_M\m{\Sigma}\m{A}_M^H} + \sigma^2.
		\end{split}
	\end{equation}
	Since $M \ge K+1$ and $\m{\Sigma}$ is positive definite, the matrix $\m{A}_M\m{\Sigma}\m{A}_M^H$ is positive semidefinite and has rank $K$. Consequently,
	\begin{equation} \label{eq:DALambdaBound}
		\begin{split}
			\lambda_K\sbra{\m{R}_{\text{DA}}}-\lambda_{K+1}\sbra{\m{R}_{\text{DA}}} = \lambda_K\sbra{\m{A}_M\m{\Sigma}\m{A}_M^H}  \ge \sigma_K^2\sbra{\m{A}_M} \lambda_K\sbra{\m{\Sigma}}  = p_{\min} \sigma_K^2\sbra{\m{A}_M}.
		\end{split}
	\end{equation}
Inserting \eqref{eq:RDAerr2} and \eqref{eq:DALambdaBound} into \eqref{eq:DASubspaceBoundApplyingDavis-Kahan} yields \eqref{eq:DASubspaceError}, completing the proof.
\end{proof}

\subsection{Signal Subspace Estimation Error of SS-ESPRIT}
Suppose that the eigen-decomposition of $\widehat{\m{R}}_{\text{DA}}$ is given by
\equ{\widehat{\m{R}}_{\text{DA}} = \sum_{j=1}^{M} \widehat{\lambda}_j \widehat{\m{u}}_j\widehat{\m{u}}_j^H = \widehat{\m{U}} \widehat{\m{\Lambda}} \widehat{\m{U}}^H + \widehat{\m{U}}_{\perp} \widehat{\m{\Lambda}}_{\perp} \widehat{\m{U}}_{\perp}^H,}
where $\lbra{\widehat{\lambda}_j }$ are sorted in descending order.
Then, it follows from \eqref{eq:RSSDA} that
\equ{\begin{split}\widehat{\m{R}}_{\text{SS}}
= \frac{1}{M}\widehat{\m{R}}_{\text{DA}}^2= \frac{1}{M}\sum_{j=1}^{M} \widehat{\lambda}_j^2 \widehat{\m{u}}_j\widehat{\m{u}}_j^H = \frac{1}{M}\widehat{\m{U}} \widehat{\m{\Lambda}}^2 \widehat{\m{U}}^H + \frac{1}{M}\widehat{\m{U}}_{\perp} \widehat{\m{\Lambda}}_{\perp}^2 \widehat{\m{U}}_{\perp}^H. \end{split}}
Therefore, the signal subspace of SS-ESPRIT is given by $\widehat{\m{U}}$, like DA-ESPRIT, if $\lbra{\widehat{\lambda}_j^2}_{j=1}^K$ are the greatest $K$ eigenvalues of $\widehat{\m{R}}_{\text{SS}}$, or equivalently, if
\equ{\widehat{\lambda}_K > \abs{\widehat{\lambda}_M}. \label{eq:lambdainequ}}
Note that the inequality in \eqref{eq:lambdainequ} might not hold when $\widehat{\lambda}_M<0$ with finitely many snapshots. In this case, DA-ESPRIT and SS-ESFPRIT are not equivalent. We have the following result.

\begin{lemma} \label{lem:SameSubpaceCondition}
	Under the assumptions of Theorem \ref{thm:DASubspaceError}, we have
	\begin{equation}
		\begin{split}
			\lambda_K\sbra{\widehat{\m{R}}_{\text{DA}}} > \abs{\lambda_{M}\sbra{\widehat{\m{R}}_{\text{DA}}}},
		\end{split}
	\end{equation}
and then SS-ESPRIT and DA-ESPRIT share the same signal subspace estimate.
\end{lemma}
\begin{proof}
Since $\m{A}_M\m{\Sigma}\m{A}_M^H$ is positive semidefinite and has rank $K$, it follows from \eqref{eq:RDA2} that
    \begin{equation} \label{eq:LambdaR_DA}
    	\lambda_K\sbra{\m{R}_{\text{DA}}} > \lambda_{K+1}\sbra{\m{R}_{\text{DA}}} = \dots = \lambda_M\sbra{\m{R}_{\text{DA}}} = \sigma^2.
    \end{equation}
Let $\m{E}_{\text{DA}} = \widehat{\m{R}}_{\text{DA}}-\m{R}_{\text{DA}}$. The assumption that the upper bound in \eqref{eq:DASubspaceError} is less than $0.586$ implies that
    \begin{equation}
    	\norm{\m{E}_{\text{DA}}}
    	\le 0.293\sbra{\lambda_K\sbra{\m{R}_{\text{DA}}}-\lambda_{K+1}\sbra{\m{R}_{\text{DA}}}}. \label{eq:EDA293}
    \end{equation}
Moreover, it follows from Theorem \ref{thm:DK} that
    \begin{equation}
    	\abs{\lambda_j\sbra{\widehat{\m{R}}_{\text{DA}}}-\lambda_j\sbra{\m{R}_{\text{DA}}}}
    	\le \norm{\m{E}_{\text{DA}}}, \quad j=1,\dots,M,
    \end{equation}
    yielding that
    \begin{equation}
    	    \lambda_K\sbra{\widehat{\m{R}}_{\text{DA}}} \ge \lambda_K\sbra{\m{R}_{\text{DA}}} - \norm{\m{E}_{\text{DA}}} \label{eq:lambdaKbd}
    \end{equation}
 and
    \begin{equation}
    	\begin{split}
    		\abs{\lambda_M\sbra{\widehat{\m{R}}_{\text{DA}}}} & = \abs{\sbra{\lambda_M\sbra{\widehat{\m{R}}_{\text{DA}}}-\lambda_M\sbra{\m{R}_{\text{DA}}}}+\lambda_M\sbra{\m{R}_{\text{DA}}}} \\
    		& \le \abs{\lambda_M\sbra{\widehat{\m{R}}_{\text{DA}}}-\lambda_M\sbra{\m{R}_{\text{DA}}}} + \abs{\lambda_M\sbra{\m{R}_{\text{DA}}}} \\
    		& \le \norm{\m{E}_{\text{DA}}} + \lambda_{K+1}\sbra{\m{R}_{\text{DA}}}.
    	\end{split} \label{eq:lambdaMbd}
    \end{equation}
    Making use of \eqref{eq:lambdaKbd}, \eqref{eq:lambdaMbd}, \eqref{eq:EDA293} and \eqref{eq:LambdaR_DA} consecutively, we have that
    \begin{equation}
    	\begin{split}
    		& \lambda_K\sbra{\widehat{\m{R}}_{\text{DA}}} - \abs{\lambda_M\sbra{\widehat{\m{R}}_{\text{DA}}}} \\
    		& \ge \lambda_K\sbra{\m{R}_{\text{DA}}}-\lambda_{K+1}\sbra{\m{R}_{\text{DA}}} - 2\norm{\m{E}_{\text{DA}}} \\
    		& \ge 0.414\sbra{\lambda_K\sbra{\m{R}_{\text{DA}}}-\lambda_{K+1}\sbra{\m{R}_{\text{DA}}}} \\
    		& > 0,
    	\end{split}
    \end{equation}
    completing the proof.
\end{proof}

Making use of Lemma \ref{lem:SameSubpaceCondition}, we have immediately the following result.
\begin{theorem} Theorem \ref{thm:DASubspaceError} remains to hold with the replacement of $\widehat{\m{R}}_{\text{DA}}$ to $\widehat{\m{R}}_{\text{SS}}$. \label{thm:SSSubspaceError}
\end{theorem}

\subsection{Error Bound for DA-ESPRIT and SS-ESPRIT}
We are ready to derive the error bound for DA-ESPRIT and SS-ESPRIT. The following theorem is a result of combining Theorem \ref{thm:DASubspaceError}, Theorem \ref{thm:SSSubspaceError} and Lemma \ref{lem:mdlem}.
\begin{theorem} \label{thm:mdBoundofDA&SSESPRIT}
	Under Assumptions A1-A3, if $M\ge K+1$, then it holds for DA-ESPRIT and SS-ESPRIT that
    \begin{equation} \label{eq:mdBoundofDA&SSESPRIT}
    	\begin{split}
 \text{md}\sbra{\widehat{\cT},\cT} \le \min \lbra{ 1 , \frac{2^{2K+9}N_{\text{S}}M\sqrt{K^3}}{p_{\min}\sigma_K^3\sbra{\m{A}_M}} \cdot \dfrac{\max\lbra{\sigma^2,p_{\max}\norm{\m{A}_{\Omega}}^2}}{\sqrt{L}} }
    	\end{split}
    \end{equation}
    with probability at least $1 - 2e^{-\frac{N_S}{2}}$.
\end{theorem}
\begin{proof}
Inserting \eqref{eq:DASubspaceError} into \eqref{eq:mdlem} yields \eqref{eq:mdBoundofDA&SSESPRIT}. Note that the unit upper bound in \eqref{eq:mdBoundofDA&SSESPRIT} holds naturally.
The conditions, $L\ge N_{\text{S}}$ and the upper bound in \eqref{eq:DASubspaceError} is less than $0.586$, in Theorem \ref{thm:DASubspaceError} are removed here since in the case when they are not satisfied, it can be shown that the upper bound in \eqref{eq:mdBoundofDA&SSESPRIT} must equal $1$.
\end{proof}

It follows from Theorem \ref{thm:mdBoundofDA&SSESPRIT} that DA-ESPRIT and SS-ESPRIT can stably estimate the frequencies if
\begin{equation}
	\frac{\max\lbra{\sigma^2,C}}{\sqrt{L}}
\end{equation}
is small, where $C$ is a problem-dependent constant that is independent of $\sigma$ and $L$. The error vanishes when $L$ approach infinity regardless of the DOAs given the source number $K\leq M-1$, which is consistent with the asymptotic performance of DA-ESPRIT and SS-ESPRIT. Therefore, there is no substantial performance gap between the practical scenario of finite $L$ and the limiting case of infinite $L$. We also note that the error does not vanish in the absence of noise, which is also consistent with existing asymptotic analyses in \cite{wang2016coarrays,liu2017cramer}. Such an error is known as the ``saturation'' error and is caused by the error of covariance estimation with finitely many snapshots.

\subsection{Resolution of DA-ESPRIT and SS-ESPRIT}
We study the resolution of DA-ESPRIT and SS-ESPRIT in this subsection. We introduce the following definition \cite{yang2023nonasymptotic}.
\begin{definition} An algorithm achieves resolution $ \Delta $ if it resolves a set of frequencies $ \mathcal{T} $, which has minimum separation $ \Delta = \min_{p \neq q}\min\lbra{\abs{f_p-f_q}, 1-\abs{f_p-f_q}} $, with precision
	\begin{equation}
		\text{md} \sbra{\widehat{\cT},\cT} < \frac{\Delta}{2}.
	\end{equation}
\end{definition}
The following result is a corollary to Theorem \ref{thm:mdBoundofDA&SSESPRIT}.
\begin{corollary} \label{cor:ResolutionOfDA&SSMUSIC}
	Under Assumption A1-A3, if $M \ge K+1$ and
    \begin{equation}
    	\begin{split}
    		& L > \frac{2^{4K+20}N_{\text{S}}^2M^2K^3}{p_{\min}^2\sigma_K^6\sbra{\m{A}_M}\Delta^2} \cdot \max\lbra{\sigma^4,p_{\max}^2\norm{\m{A}_{\Omega}}^4},
    	\end{split}
    \end{equation}
    then DA-ESPRIT and SS-ESPRIT are guaranteed to achieve resolution $\Delta$ with probability at least $1 - 2e^{-\frac{N_S}{2}}$.
\end{corollary}
\begin{proof}
	Letting the upper bounds in \eqref{eq:mdBoundofDA&SSESPRIT} be less than $\frac{\Delta}{2}$ proves the corollary.
\end{proof}

By Corollary \ref{cor:ResolutionOfDA&SSMUSIC}, the resolution of DA-ESPRIT and SS-ESPRIT can be arbitrarily high given sufficient snapshots. This is in contrast to existing nonasymptotic analyses \cite{tang2013compressed, tan2014direction,liao2014music, yang2016exact,yang2019sample, qiao2019guaranteed} for DOA estimation using SLAs in which a source separation condition is assumed, leading to a substantial resolution limit.

\section{Numerical Results} \label{sec:simulation}
In this section, we provide numerical results to validate our analysis for DA-ESPRIT and SS-ESPRIT in the case of more uncorrelated sources than sensors. In our experiments, we consider an MRA with sensor number $N_{\text{S}}=6$ and aperture $N=14$ that is given by
\begin{equation} \label{MRA}
	\Omega = \lbra{0,1,6,9,11,13}.
\end{equation}
We consider $K = 8$ uncorrelated sources with the set of frequencies
\begin{equation}
	\cT = \lbra{0.1,0.25,0.35,0.45,0.6,0.7,0.8,0.9}
\end{equation}
that corresponds to the set of DOAs $\left\{11.54^{\circ},30^{\circ},44.43^{\circ},\right.$ $\left.64.16^{\circ},-53.13^{\circ},-36.87^{\circ},-23.58^{\circ},-11.54^{\circ}\right\}$. The source signals are i.i.d.~generated from a standard complex Gaussian distribution. The array output at each snapshot is corrupted by i.i.d.~complex Gaussian noise with zero mean and variance $\sigma^2$.

\begin{figure}
	\centering  % ??????
	\subfloat{
		\includegraphics[width=5in]{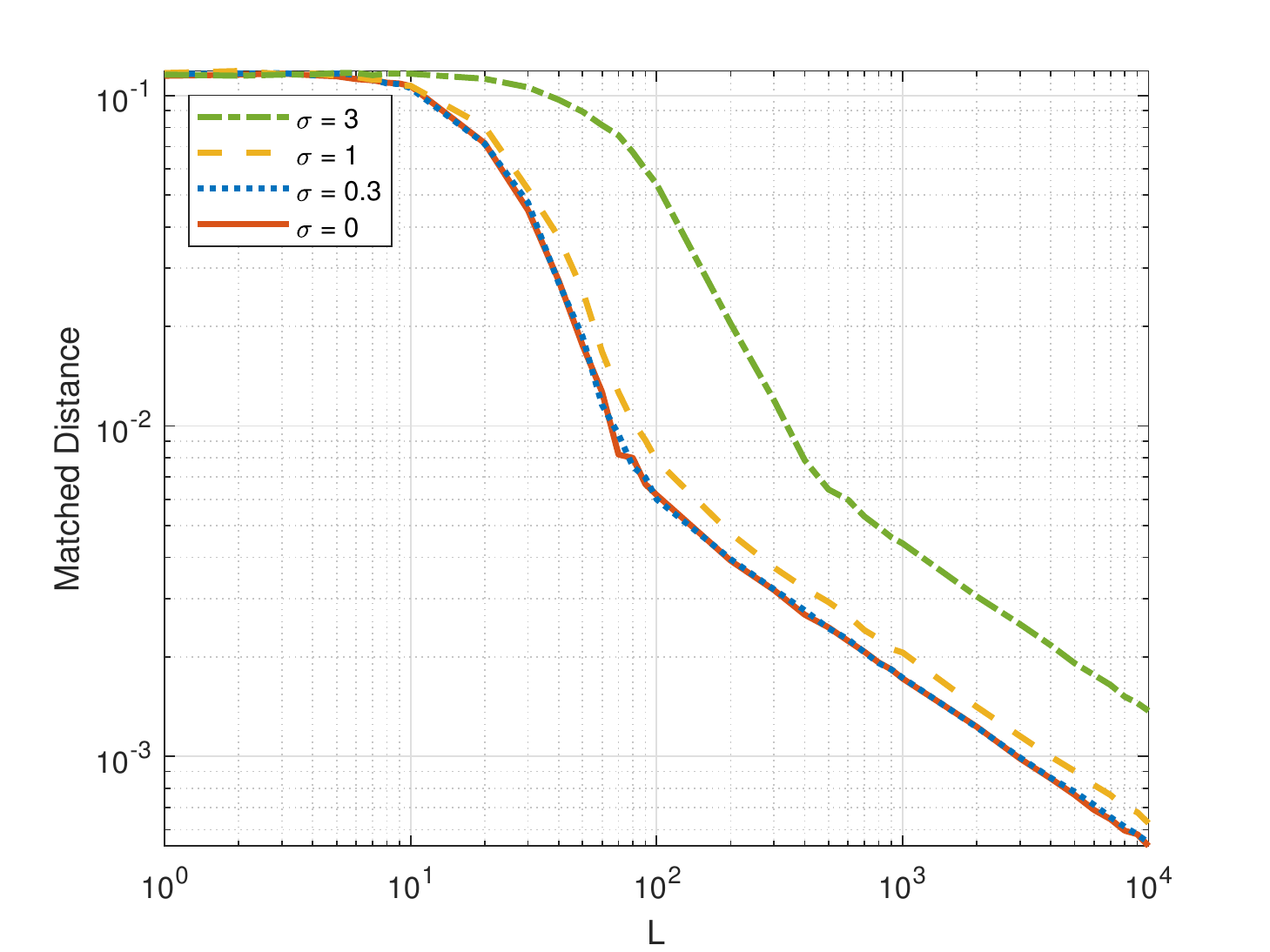}}
	\captionsetup{font={footnotesize}, labelformat=simple, labelsep=period}
	\caption{Results of matched distance of DOA estimation using DA-ESPRIT versus the snapshot number $L$.} \label{fig:2}
\end{figure}

In {\em Experiment 1}, we study the performance of DA-ESPRIT and SS-ESPRIT with varying snapshot number $L$. We consider four values of noise power with $\sigma\in\lbra{0, 0.3, 1, 3}$ and vary $L$ from $1$ to $10^4$. For each combination of $L$ and $\sigma$, we conduct 1000 Monte Carlo runs and average the results to obtain the matched distance of frequency estimation. We only present the results of DA-ESPRIT in Fig.~\ref{fig:2}, since no visible difference is shown between SS-ESPRIT and DA-ESPRIT, which is consistent with our analysis. It is seen that the curves of matched distance are approximately straight with a slope of about $-0.5$ when $L$ is large, implying that the frequency estimation error scales with $\frac{1}{\sqrt{L}}$, as predicted by Theorem \ref{thm:mdBoundofDA&SSESPRIT}. In the absence of noise, DA-ESPRIT still has an estimation error with finite snapshots, which validates Theorem \ref{thm:mdBoundofDA&SSESPRIT}.

\begin{figure}
	\centering  % ??????
	\subfloat{
		\includegraphics[width=5in]{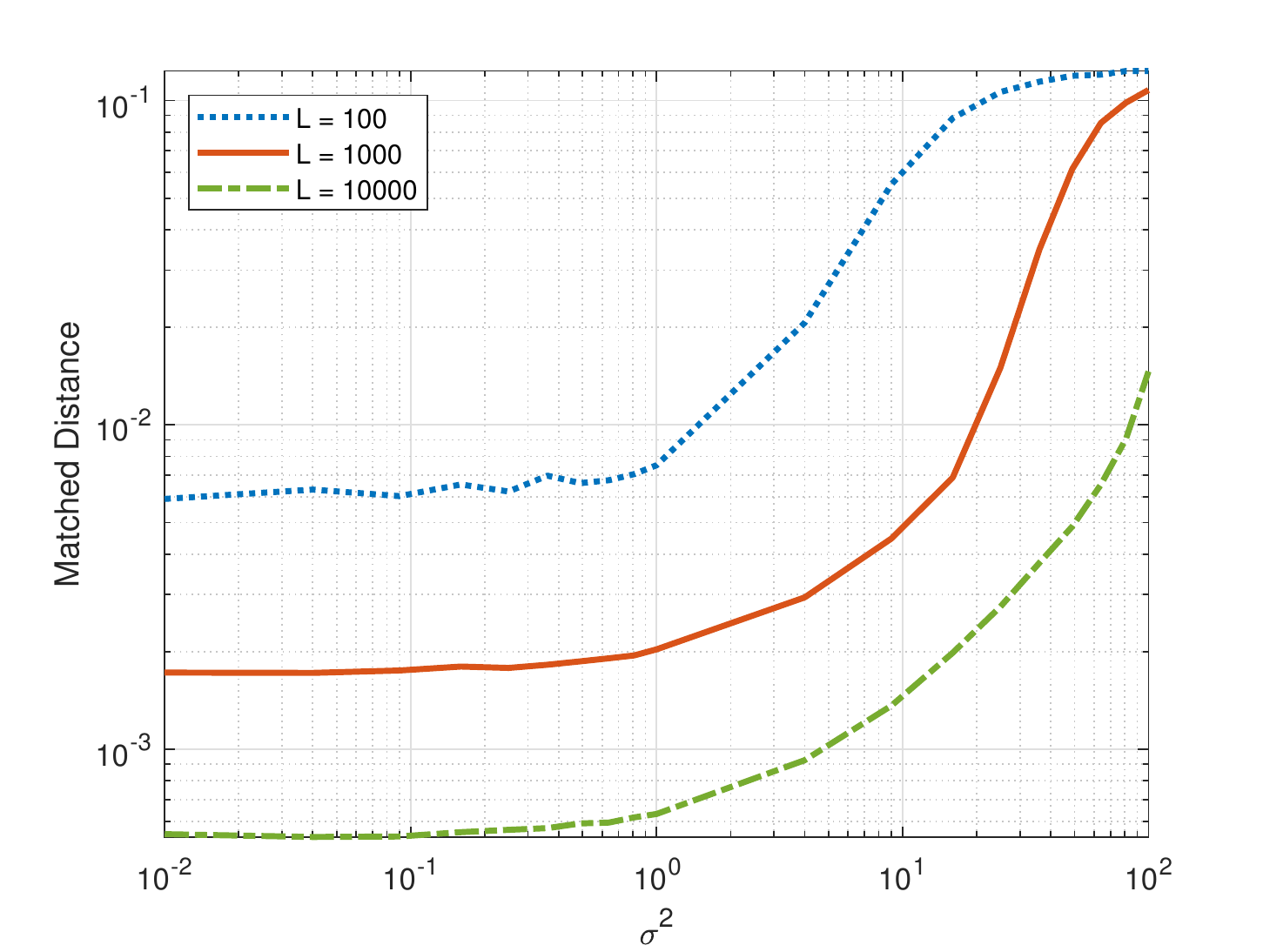}}
	\captionsetup{font={footnotesize}, labelformat=simple, labelsep=period}
	\caption{Results of matched distance of DOA estimation using DA-ESPRIT versus the noise power $\sigma^2$.} \label{fig:1}
\end{figure}

In {\em Experiment 2}, we repeat {\em Experiment 1} by letting $L = 10^2$, $10^3$ and $10^4$ and varying the noise power $\sigma^2$ from $10^{-2}$ to $10^2$. Our results of matched distance of DA-ESPRIT are presented in Fig.~\ref{fig:1}. It is seen that each curve of matched distance is approximately horizontal in the regime of small $\sigma^2$, showing again the saturation error of DA-ESPRIT. As $\sigma^2$ increases, the slope of each curve increases until when the noise level is large enough and the algorithm fails to localize the DOAs.

\begin{figure}
	\centering  % ??????
	\subfloat{
		\includegraphics[width=5in]{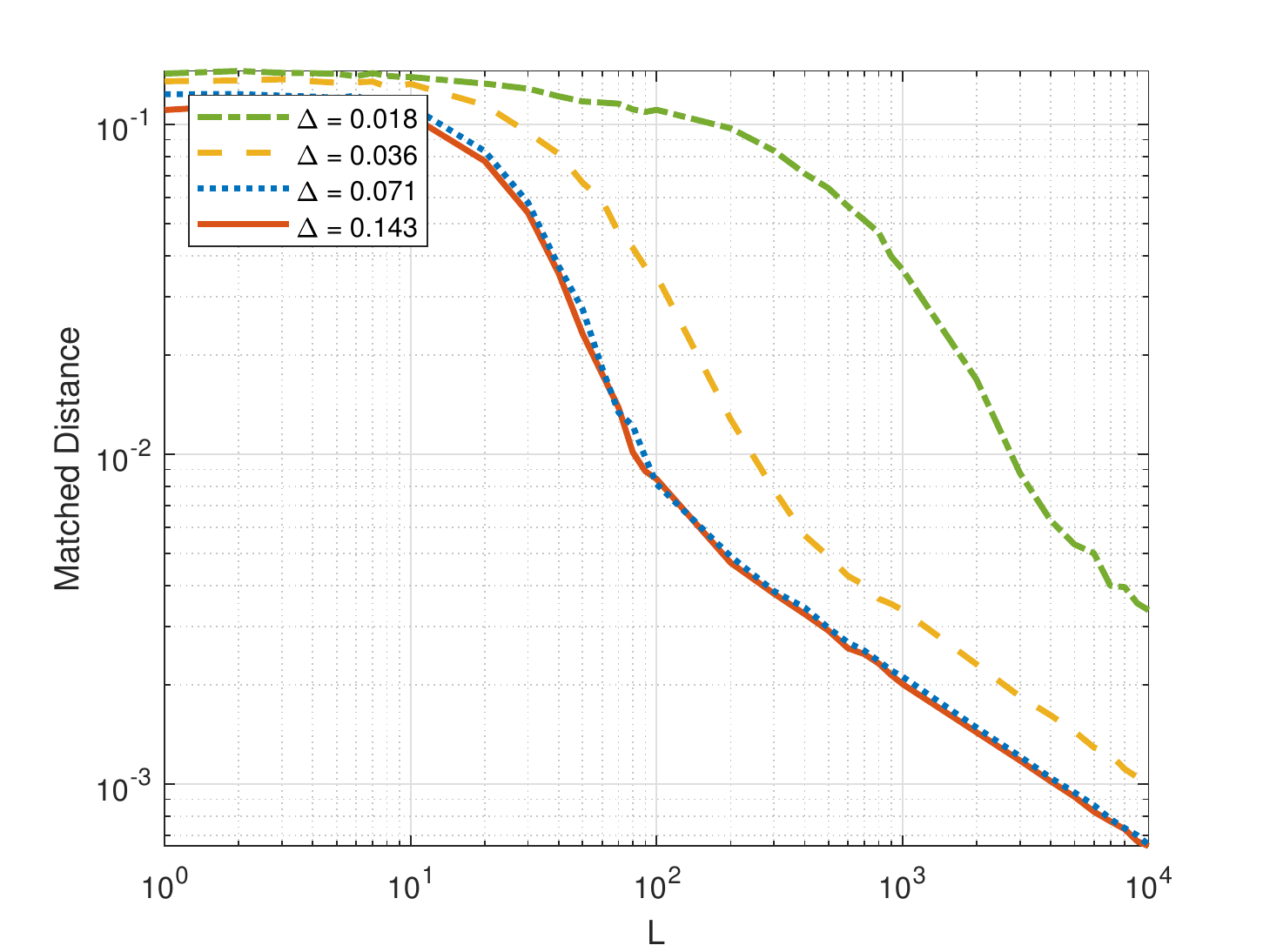}}
	\captionsetup{font={footnotesize}, labelformat=simple, labelsep=period}
	\caption{Results of matched distance of DOA estimation using DA-ESPRIT for different values of frequency separation $\Delta$ versus the snapshot number $L$.} \label{fig:3}
\end{figure}

In {\em Experiment 3}, we fix the noise level $\sigma=1$ and study the performance of DA-ESPRIT with varying frequency separation $\Delta$ and snapshot number $L$. In particular, the set of frequencies is given by
\begin{equation} \label{T}
	\cT^{\prime} = \lbra{0.1,0.25,0.35,0.45,0.6,0.7,0.8,0.8+\Delta},
\end{equation}
where $\Delta\in\lbra{0.018,0.036,0.071,0.143}$. Note that the case of $\Delta < \frac{1}{N} = 0.071$ is usually referred to as the super-resolution regime. Our results are presented in Fig.~\ref{fig:3}. It is seen a smaller frequency separation $\Delta$ leads to a larger estimation error of DA-ESPRIT, especially in the super-resolution regime. For all values of $\Delta$, DA-ESPRIT has an error with the same scaling behavior with respect to the snapshot number, as predicted by Theorem \ref{thm:mdBoundofDA&SSESPRIT}. Moreover, the resolution of DA-ESPRIT increases constantly as $L$ increases, which validates Corollary \ref{cor:ResolutionOfDA&SSMUSIC}.

%\begin{figure}
%	\centering  % ??????
%	\subfloat{
%		\includegraphics[width=\columnwidth]{EXP3-eps-converted-to.pdf}}
%	\captionsetup{font={footnotesize}, labelformat=simple, labelsep=period}
%	\caption{Results of matched distance of DOA estimation for different frequency separation $\Delta$ versus the noise power $\sigma^2$ using DA-ESPRIT.} \label{fig:4}
%\end{figure}
%
%In summary, the numerical results show that the scaling laws of the estimation error of DA- and SS-ESPRIT with respect to the noise level $\sigma$ and the snapshot number $L$. It is also seen that DA- and SS-ESPRIT can achieve arbitrary revolution providing sufficiently large $L$.

\section{Conclusion} \label{sec:conclusion}
In this paper, we derived the nonasymptotic error bounds of DA-ESPRIT and SS-ESPRIT for DOA estimation using SLAs. It is shown that DA-ESPRIT and SS-ESPRIT can stably localize more uncorrelated sources than sensors with overwhelming probability in the practical scenario of finite snapshots. They have a higher resolution as the snapshot number increases and do not suffer from a substantial resolution limit. This reveals great potentials of DA-ESPRIT and SS-ESPRIT, especially in the presence of a large number of snapshots.

%In this paper, we derived the nonasymptotic error bound of DA- and SS-ESPRIT. The results show that DA- and SS-ESPRIT can stably estimate the DOAs with overwhelming probability in the practical scenario with finite snapshots and finite SNR, and also imply that DA- and SS-ESPRIT can achieve arbitrary resolution as soon as the snapshots number is large enough.

%
%% \balance                      % 双栏
%%\nocite{*}                    % 显示所有文献
%\bibliographystyle{IEEEtran}  % 格式
%%\bibliography{References}     % bib文件名
%%\bibliography{D:/OneDrive/Research/MyWork/myreferences1}
%\bibliography{D:/OneDrive/Research/MyWork/!!!2021-SS-MUSIC-开杰/myreferences1}


\end{document}